\documentstyle[12pt,fleqn]{article}

\title{Experimental Conditions for the \\
Gamma Optical Scattering\\
September 1979\footnote{Typed in 1999 after the original September 1979 
manuscript}}

\author{Silviu Olariu
\thanks{Present address:
Institute of Physics and Nuclear Engineering,
Department of Fundamental Experimental Physics,
76900 Magurele, P.O. Box MG-6, Bucharest, Romania; 
e-mail: olariu@ifin.nipne.ro}}

\begin{document}
\date{}
\maketitle

The interaction of gamma-ray photons with a nucleus, mediated by an
electromagnetic field, was investigated in our paper, The Tuning of
$\gamma$-Ray Processes with High Power Optical 
Radiation.\footnote{Note added in May 1999: the referenced work
is S. Olariu et al., Phys. Rev. C {\bf 23}, 50 (1981); it was submitted 
for publication on 30 November 1979.} 
In that paper little attention was 
given to the absolute location of the energy levels of the nuclei: 
it was rather a discussion of the resonance of the gamma ray energy 
and of the electromagnetic energy to the nuclear transition energy, 
where the energy of the gamma ray was considered as a parameter. 

In fact, there is a close relationship between the energy of the gamma ray from
the nucleus in the source and the transition energy of the nucleus in the
absorber. The positions of the energy levels of a free nucleus are affected by
the interaction with the electric and magnetic fields in which the nucleus is
immersed.\footnote{We follow in this paragraph G. K. Shenoy and F. E. Wagner,
Editors, {M\"{o}ssbauer Isomer Shifts}, North-Holland, 1978, 
The Introduction.}  These fields can be created by external charge
distributions or by atomic electrons. For example, the magnetic dipole
interaction arises if an inner unpaired electron polarises the electron
configuration, so that there is a difference in the energies corresponding to
the various orientations of the nuclear magnetic moment, which is of the order
of $10^{-5}$ eV. The electric quadrupole interaction appears for ellipsoidal
nuclei, which have a dependence of their energy on the orientation in the
spatially varying electric fields created by a non-spherical charge
distribution.  This interaction is again in the order of $10^{-5}$ eV. The
isomeric monopole interaction arises from the dependence of the interaction
energy between the external electronic charge and the nucleus on the nuclear
radius, which is different for the ground and the excited state. The isomeric
energy of interaction is of about $10^{-7}$ eV. 

The hyperfine interaction mentioned above produce either splittings, or shifts,
of the levels. The binding of the nuclei in a lattice eliminates the Doppler
and the recoil effects, but, as the chemical composition of the source and of
the absorber is not generally the same, the internal fields will be different,
and the emission and absorption lines will not overlap. We proposed in our
previous papers to compensate this detuning by an applied electromagnetic field
of appropriate energy. Since 1 eV corresponds to $2.41\times 10^{14}$ Hz, the
magnetic dipole and electric quadrupole spilttings are compensated by fields of
frequencies in the range of several GHz, while the isomeric monopole shift is
compensated by frequencies in the range of 10 Mz. 

As a matter of fact, the magnetic hyperfine interaction can be seen directly by
Nuclear Magnetic Resonance techniques, the electric quadrupole interaction can
be measured by Nuclear Quadrupole Resonance techniques, and the monopole
interaction usually called the isomer shift is specific to the M\"{o}ssbauer
technique. Of course, each of these have their own limits of applicability, and
a new technique may be efficient where the existing ones are not. But the main
interest in the Gamma-Optical technique is to obtain experimental evidence for
the nuclear Raman scatterings which represents a step toward tunable sources of
$\gamma$ radiation, and a possible way to the Gamma RayLaser.

The results of our paper, The Tuning of the $\gamma$-Ray Processes \dots , was
that the ratio $B$ of the cross section of the enhanced scattering,
$\sigma^{(2)}$, to the single-photon, Breit-Wigner, cross section,
$\sigma^{(1)}$, is, for transitions mediated by the magnetic sublevels,
proportional to the density of the energy flux, $\Phi_2$, of the
electromagnetic field, divided by the square of the frequency $\omega_2$,
\begin{equation}
B\equiv \sigma^{(2)}/\sigma^{(1)}=\frac{4\pi e^2 \Phi_2 g^2}{m^2 c^3
\omega_2^2} .
\end{equation} 
It was also shown that the ratio, $Z$, of the gamma-optical cross section,
$\sigma^{(2)}$, to the off-resonance, single-photon cross section,
$\sigma_{\pm\omega_2}^{(1)}$, is proportional to the energy flux, $\Phi_2$,
divided by the square of the width, $\Gamma$, of the level,
\begin{equation}
Z\equiv\sigma^{(2)}/\sigma_{\pm\omega_2}^{(1)}=\frac{4\pi e^2 \Phi_2
g^2}{m^2c^3\Gamma^2} .
\end{equation}
The dependence $1/\omega_2^2$ of the two-photon cross section is valid with
unsplit, Lorentzian profiles of the lines. The preparation of samples which
provide such lines is certainly possible, by using lattices of appropriate
chemical composition and internal symmetry. On the other hand, the internal
fields, or applied external fields, produce splittings of the energy levels,
and the $1/\omega_2^2$ dependence of the two-photon cross section is valid only
for energies $\hbar \omega_2$ larger than the splitting. Below that limit, the
cross section becomes independent of the frequency $\omega_2$ of the
electromagnetic field. 

The $\Phi_2$ dependence of the gamma-optical cross section is valid if the
power broadening of the lines is smaller, or comparable to the effective
linewidth. The Zeeman shift of the levels becomes comparable to the width of
the lines at intensities of the magnetic field of about 100
Gauss.\footnote{ibid., p. 567, Fig. 8d.1} At very large powers, we expect to
see a saturation of the cross section, together with the modifications
predicted in Ref. 15 of our third Gamma-Optical paper.\footnote{Note added 
in May 1999: this reference is M. N. Hack and M. Hamermesh, Nuovo Cimento 
{\bf XIX}, 546 (1961).}

Due to imperfect preparation of the samples, the cross section of the two 
photon process could be lower than that predicted in Eq. 1 by the ratio of the
natural linewith, $\Gamma$, to the effective linewidth, $\Gamma_e$. On the
other hand, the ratio of the cross sections $\sigma^{(2)}$ and $\sigma^{(1)}$
is determined by the natural linewidth, as written in Eq. 2, because the
single-photon, Breit-Wigner cross section, is also reduced by the same ratio
$\Gamma/\Gamma_e$. 

A low frequency $\omega_2$ results in lower power requirement $\Phi_2$ for a
given branching ratio $B$, as may be seen from Eq. 1, and a narrow linewidth
$\Gamma$ results in lower power requirement for a given signal to noise ratio
$Z$, as may be seen from Eq. 2. 

We have seen that the types of hyperfine interactions create two ranges of
frequencies $\omega_2$. Beyond 1 GHz, which corresponds to a wavelength of 30
cm, in the region of the dipole and quadrupole interaction, the high values of
the power density $\Phi_2$ are most conveniently obtained by the
microwave-cavity technique.

This technique is not appropriate at the lower frequencies in the range of 10
MHz, because the size of the cavity, which is determined by the wavelength of
the oscillating field, becomes impracticable. But, going back to the bases of
the interaction between the gamma ray photon and the nucleus, mediated by an
external applied field, we see that all that is necessary for the interaction
mediated by the magnetic sublevels is the presence of a magnetic field $H_2$,
oscillating with the frequency $\omega_2$. Currents oscillating in conductors
create such magnetic fields, which are not electromagnetic waves. We have in
fact an LC circuit, and, if we assume the volume of the capacitor and of the 
inductor to be comparable, then the amplitude of the magnetic field in the
inductor is comparable (in CGS units) to the amplitude of the electric field in
the capacitor. A magnetic field of 100 Gauss corresponds to an electric field
of 100 statvolt/cm, or 30 kV/cm. The energy periodically transferred between
the inductor and capacitor is $H^2/8\pi$ erg/cm$^3$. Assuming $H$=100 Gauss, a
volume of $10^2$ cm$^3$, and a $Q$-factor of the LC circuit of 100, the input
power in the circuit is, at a frequency of 100 MHz, of about 1 kW. The
equivalent energy flux $\Phi_2$ corresponding to the field $H_2$ of 100 Gauss
is of about $10^6$ W/cm$^2$.

For illustrative purposes we chose here the 6.2 keV, 6.8 $\mu$sec line of
$^{181}$Ta, which has a single-photon cross section of $1.73\times 10^{-18}$
cm$^2$; $^{181}$Ta has a good natural abundance (99.99 \%), the 6.2 keV level
is populated by the decay of $^{181}$W, which has a half-life of 140 days and
is obtained by neutron irradiation of $^{180}$W. The branching ratio $B$
corresponding to a field $H_2$ of 100 Gauss and a frequency $\omega_2$ of 10
MHz is $B\approx 10^{-3}$. The cross section of the two-photon process expected
under these conditions is $\sigma^{(2)}\approx 10^{-22}$ cm$^2$, because of the
broadening of the lines due to imperfect sample preparation and to the power
broadening. The ratio $Z=\sigma^{(2)}/\sigma_{\pm \omega_2}^{(1)}$ is under 
these conditions, $Z\approx 10$.

The expression for the branching ratio $B$ and the signal to noise ratio $Z$
were derived by assuming Lorentzian profiles for the two modes of the
electromagnetic field. If the sampples are immersed in an oscillating magnetic
field, it can be seen from the basic equations for the amplitude of a
two-photon process that, if the transition is mediated by the magnetic
sublevels, the branching ratio is
\begin{equation}
B=\left(\frac{g\frac{e\hbar}{2m c}H_2}{\hbar \omega_2}\right)^2 ,
\end{equation}
which is similar to Eq. 1, since the equivalent energy flux is
$\Phi_2=\frac{c}{8\pi}H_2^2.$

The relatively high values of the branching ratio $B$ and the good values 
of the signal to noise ratio $Z$ obtained with a relatively low imput power
which may be tuned by varying the capacitance in the LC circuit suggest that
the experimental approach to the gamma-optical absorption be focused on
processes compensated by smaller frequencies, like the isomeric shift and the
second order Doppler shift, and involving narrow lines.

We have to obtain first the resonance when the frequency of the oscillating
magnetic field corresponds to the isomeric shift between the source and
absorber. Then, if a static uniform magnetic field is superposed to the
oscillating magnetic field, we expect to see a plateau of the two photon cross
section at lower values of $H_2$, and a decrease of the cross section as
$1/H^2$ as the intensity of the static field is increased, corresponding to the
fact that the Zeeman splitting becomes larger than the value of the frequency
$\omega_2$ determined by the isomeric shift. Finally, without the static field,
increasing the intensity of the oscillating field, we expect to obtain the
saturation of the cross section and to see the changes in the shape of the line
described in Ref. 15 of the third Gamma-Optical paper.

There is, of course, complete symmetry between compensating the detuning in
source or absorber. Yet we suggest that the {\it source} be immersed in the
oscillating field, in order to obtain the experimental confirmation of the
nuclear Raman scattering, which is more close to a tunable gamma ray source,
and is somewhat related to a gamma-ray laser. 

Due to the relatively large values of the branching ratio, $B$, - a 
$B=10^{-4}$ is probably not difficult to obtain -, apparently there is no
problem with the counting rate, so that weak radioactive source, say 1 mCi, can
be used (1 Ci=$3\times 10^{10}$ disintegrations/sec). The absolute value of the
two photon cross section is in the range of $10^{-23}$ cm$^2$, and is
comparable to other scattering processes, like the Compton scattering and the
photoelectric effect. But we may take the advantage of the fact the
fluorescence radiation has a definite energy, while the Compton scattered
radiation has a continuous spectrum, and to discriminate energetically between
the signal and the noise. The resolution of the gamma-ray detectors in the
region of interest in M\"{o}ssbauer spectroscopy is good, it seems to be
$\approx 100$ eV. For the single-photon off-resonance scattering, the ratio $Z$
is higher than 1 at larger magnetic intensities.\footnote{The intensity is
limited by the condition that the power broadening be comparable to the natural
width.} At lower intensities where $Z<1$ other discrimination techniques are
available, such as the comparing of the signals when the magnetic power is off,
and on.

Since the frequency $\omega_2$ is in the range 10 MHz the width of the line has
to be in the range 1 $\mu$sec. There are three narrow lines in M\"{o}ssbauer
spectroscopy, namely \\
$^{67}$Zn, 93.3 keV, 9.2 $\mu$sec, $4.93\times 10^{-20}$ cm$^2$,\\
$^{73}$Ge, 13.3 keV, 3.9 $\mu$sec, $0.76\times 10^{-20}$ cm$^2$,\\
$^{181}$Ta, 6.2 keV, 6.8 $\mu$sec, $1.73\times 10^{-18}$cm$^2$.\\
The best candidate seems to be $^{181}$Ta. In this case, an isomeric shift of 1
mm/sec is equivalent to $2.07\times 10^{-8}$ eV, or 5.0 MHz. Many useful
informations on the spectroscopy of the 6.2 keV gamma ray of $^{181}$Ta and
related isomeric shifts can be found in the cited reference.\footnote{ibid, pp.
563-591, p. 873, p. 879, p. 886}

In conclusion, the gamma-optical experiment requires an LC circuit operating in
the range of 10 MHz, with an inductor providing magnetic fields of the order of
10 Gauss, the circuit being tunable by the variation of the capacitance C; the
narrow-line sources and absorbers; and conventional gamma-ray equipment. Since
the frequency of 10 MHz is in the radio frequency range, difficulties could
arise only from the magnitude of the field. Seemingly, a magnetic field of 1
Gauss, which is equivalent to 300 V/cm, is sufficient for the experiment.

\end{document}